\documentclass[letterpaper, 10 pt, conference]{ieeeconf}  

\IEEEoverridecommandlockouts                              
\usepackage{times}
\usepackage[utf8]{inputenc}
\usepackage{xcolor}
\usepackage{amsmath}

\usepackage{amsthm}
\usepackage{amsfonts}
\usepackage{amssymb}
\usepackage{graphicx}
\usepackage{float}
\usepackage[justification=centering,font=small,labelfont=bf]{caption}
\usepackage{tikz}
\usepackage[noadjust]{cite}
\usepackage{hyperref}
\usepackage{comment}
\usepackage{tabularx}
\usepackage{algorithm}
\usepackage{algpseudocodex}

\newcommand{\R}{\mathbb{R}}

\DeclareMathAlphabet{\mymathbb}{U}{BOONDOX-ds}{m}{n}

\newcommand{\W}{\mathcal{W}}
\newcommand{\K}{\mathcal{K}}

\newcommand{\of}{\circ}

\newcommand{\id}{\mathcal{I}}

\theoremstyle{plain}
\newtheorem{thm}{Theorem}
\newtheorem{lem}[thm]{Lemma}

\theoremstyle{definition}

\theoremstyle{remark}

\newtheorem*{sktch}{Sketch of Proof}

\newtheorem{problem}[thm]{\bf Problem}

\newcommand\copyrighttext{%
\footnotesize\textcopyright 2024 IEEE. Personal use of this material is permitted.  Permission from IEEE must be obtained for all other uses, in any current or future media, including reprinting/republishing this material for advertising or promotional purposes, creating new collective works, for resale or redistribution to servers or lists, or reuse of any copyrighted component of this work in other works.}
\newcommand\copyrightnotice{
\begin{tikzpicture}[remember picture,overlay]
\node[anchor=south,yshift=10pt] at (current page.south) {\fbox{\parbox{\dimexpr\textwidth-\fboxsep-\fboxrule\relax}{\copyrighttext}}};
\end{tikzpicture}}

\title{\LARGE \bf
Causal Tracking of Distributions in Wasserstein Space: \\ A Model Predictive Control Scheme 
}

\author{Max Emerick, Jared Jonas, and Bassam Bamieh
\thanks{M. Emerick, J. Jonas, and B. Bamieh are with the Department of Mechanical Engineering,
		University of California, Santa Barbara, USA
        {\tt\small \{memerick,jjonas,bamieh\}@ucsb.edu}}%
}

\begin{document}

\maketitle
\thispagestyle{empty}
\pagestyle{empty}

\begin{abstract}
We consider a problem of optimal swarm tracking which can be formulated as a tracking problem for distributions in the Wasserstein space. Optimal solutions to this problem are non-causal and require knowing the time-trajectory of the reference distribution in advance. We propose a scheme where these non-causal solutions can be used together with a predictive model for the reference to achieve causal tracking of a priori-unknown references. We develop a model-predictive control scheme built around the simple case where the reference is constant-in-time. A computational algorithm based on particle methods and discrete optimal mass transport is presented, and numerical simulations are provided for various classes of reference signals. The results demonstrate that the proposed control algorithm achieves reasonable performance even when using simple predictive models.
\end{abstract}

\copyrightnotice
\vspace{-1em}

\section{Introduction}

Recent advances in processing and communication hardware have made it possible to start using swarms of autonomous mobile robots to accomplish tasks in diverse settings. The potential application areas of these swarms are vast, including emergency response, environmental monitoring, transportation, logistics, data collection, and defense. In many of these application areas, it is desirable to use large swarms due to their efficiency and robustness. However, as swarms scale in size, it also becomes increasingly difficult to plan and coordinate motion. For very large swarms of interacting agents, even simulating the behavior of these swarms can become intractable.

One approach to understand performance limitation issues in these problems is to model large-scale swarms as distributions (i.e. as continua) instead of as collections of individual agents. This effectively discards ``microscopic'' information pertaining to the states of individual agents while retaining ``macroscopic'' information pertaining to the state of the overall swarm. This provides both a significant model reduction and a scale-independent approach to analyzing swarm behavior. Thus, the problem of controlling distributions is a problem of interest\footnote{The problem of controlling distributions also arises in probabilistic settings where only a probability distribution of possible system states is available, and in ensemble systems, where a large collection of identical subsystems is described in terms of the distribution of subsystem states.}.

Much of the literature surrounding control of distributions has focused on the problem of state transfer, where the goal is to drive the state of the system from some initial distribution $A$ to a final distribution $B$. The classic problem of coverage control, for instance, can be formulated as a state transfer problem. In this paper, we focus on the problem of tracking instead, where the state distribution needs to continuously track a moving reference. Many desirable swarm behaviors can be expressed in terms of tracking.

Compared to the literature on state transfer for distributions, the literature on tracking is considerably lighter. In \cite{Halder2014}, a problem is investigated where a series of system models is generated to interpolate between successive observations of ensemble states. In \cite{Foderaro2016}, a distributed numerical algorithm is developed to optimize tracking of stochastically moving targets in large-scale sensor networks. In \cite{Chen2018}, a numerical method is developed to smoothly interpolate between a sequence of probability distributions. In \cite{Ren2019}, the literature on probability density function (PDF) control is reviewed, where the objective is to control a stochastic system by shaping the PDF of its output. In \cite{Zhou2021}, a neural network is used to approximate an optimal control for a swarm to track a provided reference. In \cite{Abdulghafoor2023}, a motion control algorithm for multi-agent systems is proposed where distributed tracking control comprises one step in the overall algorithm.

In this present paper, we propose a model-predictive control (MPC) algorithm for distribution tracking which is based on optimal mass transport. This algorithm builds on our previous work on this problem \cite{Emerick2022,Emerick2023}. Tools from optimal mass transport (and the Wasserstein distance in particular) have gained recent popularity in distribution control problems (see, e.g., \cite{Chen2021} and references therein). Our approach differs from many of these approaches in that our model is motivated largely by applications in swarm control, we treat the problem of tracking (as opposed to state transfer), and our algorithm handles general (i.e. non-Gaussian) densities and can be implemented in real time.

The rest of this paper proceeds as follows. In Section \ref{background}, we present the necessary background. In Section \ref{mpc_scheme}, the MPC algorithm is developed. In Section \ref{simulations}, we provide simulation results. Section \ref{conclusion} concludes with a brief summary and a discussion of future work.

\section{Background and Problem Formulation} \label{background}

In this section, we present the background necessary to motivate and explain our MPC scheme.

\subsection{Notation and Preliminaries}

This paper relies heavily on several concepts from optimal mass transport including the Monge and Kantorovich problems, transport maps and plans, the Wasserstein distance, and Wasserstein geodesics. For a short and readable introduction to these topics, the authors recommend \cite{Santambrogio2010}.

We will refer to the entities which we are trying to track and control as ``distributions'' (as in ``mass distribution'' or ``probability distribution''). From a measure-theoretic point of view, these distributions are density functions of finite measures. Thus, ``distribution'' is used more or less synonymously with ``measure'' or ``density'' in this paper.

Throughout this paper, the notations $f$ or $f(\cdot)$ are used interchangeably and refer to a whole function as an object, while the notation $f(t)$ refers to the specific value that function takes at time $t$. We also use subscripts to denote parameters; for example, $R$ will refer to a parameterized curve in the space of distributions, while $R_t$ will refer to a specific distribution at time $t$, and $R_t(x)$ will refer to the value of that specific distribution at location $x$.

\subsection{Problem Setting and Motivation}

We briefly describe the  problem setting here. We refer to~\cite{Emerick2022} for the full development and motivation.
In our setting, there are two entities, referred to as the ``resource'' and ``demand''. The resource represents a swarm of controlled mobile agents, and the demand represents some entity which the resource aims to track. The reason for this terminology is that this latter entity ``demands'' some supplies or services which the resource provides. Both the resource and demand are modeled by time-varying distributions over the domain. The resource and demand distributions at time $t$ are denoted by $R_t$ and $D_t$ respectively, and the domain $\Omega$ is assumed to be a compact convex subset of $\R^n$.

The dynamics of the resource are given by the \emph{transport equation} (also called the \emph{advection} or \emph{continuity} equation)
\begin{equation}
	\partial_t R_t(x) ~=~ - \nabla \cdot ( V_t(x) \, R_t(x) ) ,
\end{equation}
which describes the evolution of the distribution $R_t$ as it is transported by the velocity field $V_t$ under the conservation of mass. In our setting, the velocity field $V_t$ is considered to be a control input.

The demand $D$ is assumed to be an external signal that the resource $R$ aims to track.
We  set up an optimal tracking control problem which trades off between two competing objectives. The first objective, $\W_2^2 ( R_t,D_t )$, denotes the square of the \emph{2-Wasserstein distance} between the resource and demand. This distance function penalizes how far the resource is from meeting the demand. In the optimal transport literature, the Wasserstein distance is usually motivated as the cost of moving one distribution to another, e.g., as in the term ``earth mover's distance''. We emphasize that in the present setting, the Wasserstein distance is not a motion cost, but rather the cost of servicing or assignment between the resource and demand distributions. We account for motion cost  of the resources differently as described next.   
	
\begin{figure}[!ht]
	\centering
	\includegraphics[width=0.95\linewidth]{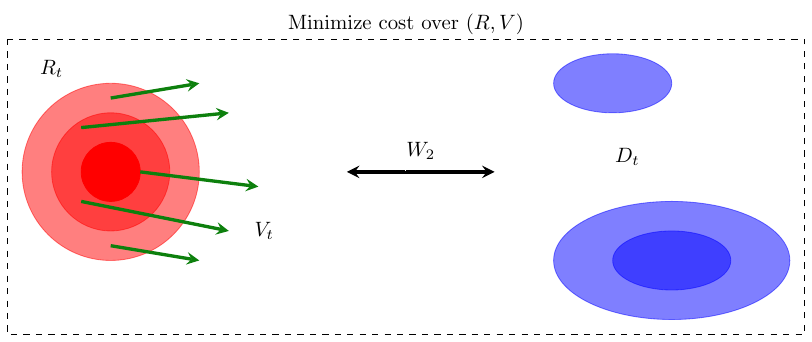}
	\caption{Pictorial representation of Problem \ref{orig_prob}. The resource $R$ aims to track the demand $D$ via transport by the velocity field $V$. The total cost is minimized over the maneuver.}
	\label{model_drawing}
\end{figure}

The second objective, $\| V_t \|_{L^2(R_t)}^2$, denotes the square of the \emph{$R_t$-weighted $L^2$ norm} of $V_t$
\begin{equation}
	\| V_t \|_{L^2(R_t)}^2 ~:=~ \int_\Omega \Vert V_t(x) \Vert_2^2 \, R_t(x) \, dx .
\end{equation}
This norm penalizes how aggressively the resource moves. Thus, this optimal tracking control problem trades off between the competing costs of distance-to-demand and efficiency-of-movement. In other words, $R_t$ should track $D_t$, but in an ``efficient'' way. This notion of efficiency is made precise by the total objective function
\begin{equation}
	\int_0^T \W_2^2 ( R_t,D_t ) + \alpha^2 \| V_t \|_{L^2(R_t)}^2 \, dt ,
\end{equation}
which integrates these two costs over the time horizon $[0,T]$. The parameter $\alpha$ in the above equation is a constant ``trade-off parameter'' which sets the relative importance of these two costs -- a small value of $\alpha$ means that motion is cheap and results in an aggressive control which tracks the demand closely, while a large value of $\alpha$ means that motion is expensive and results in a conservative control which prioritizes efficiency of movement. All in all, the problem is written formally as follows.
\begin{problem} \label{orig_prob}
	Given an initial resource distribution $R_0$ and demand signal $D$, solve
	\begin{equation} \label{original}
		\begin{split}
			\inf_{R,V} & \, \int_0^T \W_2^2 ( R_t,D_t ) + \alpha^2 \| V_t \|_{L^2(R_t)}^2 \, dt \\
			&\text{s.t.} \quad \partial_t R_t ~=~ - \nabla \cdot ( V_t R_t ) .
		\end{split}
	\end{equation}
\end{problem}
We consider this problem as a nonlinear, infinite-dimensional optimal tracking control problem, with $R$, $D$, $V$, $\W_2^2(R,D)$, and $\| V \|_{L^2(R)}^2$ considered as the state, reference, control input, tracking error, and control effort, respectively. As is common in optimal control, we will often be interested in solutions in feedback form.
Figure \ref{model_drawing} shows a pictorial representation of this model.

In previous works \cite{Emerick2022,Emerick2023}, we found explicit solutions to this problem in the special cases where the spatial domain is one-dimensional (i.e. $\Omega \subset \R$), and where the demand $D$ is static (i.e. constant-in-time). In those works, it was shown that solutions to this problem yield controllers which are \emph{noncausal}. In other words, the optimal control at time $t$ depends on the future of the demand $D$. While these solutions lend insight into the optimal system behavior and may be implemented if a good prediction of the future demand is available, they cannot be used for real-time tracking control for demands which are unknown a priori. This forms the motivation for the current work. In this paper, we treat this issue of causality by developing an MPC algorithm which uses the noncausal optimal solutions together with a predictive demand model to attain causal real-time tracking control of a priori-unknown demands.

\subsection{Structure of Noncausal Solutions}

Before presenting our proposed MPC scheme, we first describe the noncausal structure of solutions to the original problem \eqref{orig_prob}. The necessary conditions for optimality for this problem can be written in the following form:
\begin{align}
	\partial_t R_t ~&=~ - \nabla \cdot (V_t \, R_t) \label{necc_cond_3} \\
	\partial_t V_t ~&=~ - {\textstyle \frac{1}{2}} \nabla \| V_t \|^2 - {\textstyle \frac{1}{\alpha^2}} (M_{R_t \to D_t} - \id) . \label{necc_cond_4}
\end{align}
Here, \eqref{necc_cond_3} is the state equation, describing the evolution of $R_t$ under the control velocity field $V_t$, while \eqref{necc_cond_4} describes the evolution of the optimal control $V_t$ under the forcing term $(M_{R_t \to D_t} - \id)$. Here, $M_{R_t \to D_t}$ is the optimal transport map (i.e. Monge map) taking $R_t$ to $D_t$ and $\id$ is the identity map on $\Omega$. These necessary conditions form a two-point boundary value PDE with $R_t = R_0$ specified at $t = 0$ and $V_T = 0$ specified at $t=T$. Because these equations are coupled, they need to be solved simultaneously, and their solutions depend on the entire trajectory of $D$, which enters through the forcing term. Thus the entire trajectory of $D$ must be known beforehand in order to solve these equations.

We point out that there are cases where this noncausality is acceptable, namely, in cases where the demand evolves deterministically under known dynamics or is predetermined by a designer. For example, if the demand is known to be static or periodically time-varying, then we can anticipate the entire future of $D$. Similarly, in an application such as a drone lightshow where the sequence of formations and the paths that individual drones take are specified beforehand, the entire future of $D$ can be made available to the controller as well. However, this noncausality prevents these solutions from being applied in tracking scenarios where the demand is unknown a priori, and this motivates the development of the MPC scheme as described next.

\section{Model-Predictive Control Scheme} \label{mpc_scheme}

In this section, we develop and explain our proposed MPC scheme. The central idea here is the same as any MPC scheme: iteratively solve an optimal control problem and apply only the first step of the optimal control before repeating the process. The optimal control problem that we will solve will be the infinite-horizon counterpart of the original (noncausal) problem \eqref{original}, and we will handle the noncausality of optimal solutions to this problem by using a \emph{prediction} of the demand trajectory in place of the actual trajectory. In other words, this MPC scheme is a way to create a causal controller out of a noncausal controller and a predictive model. While it seems unlikely that this is a new idea, we were unable to find any references presenting exactly this concept in this way.

In further detail, this MPC scheme has four steps:
\begin{enumerate}
	\item Use a predictive model to forecast the demand trajectory from the current state $D_t$.
	\item Use this predicted trajectory $\hat{D}$ in the necessary conditions for optimality \eqref{necc_cond_3} - \eqref{necc_cond_4} to find a control $V_t$.
	\item Apply the computed control $V_t$ over a short time horizon of length $\Delta_t$.
	\item Update the state of the demand $D_t$ and repeat.
\end{enumerate}

\begin{figure}[!ht]
	\centering
	\includegraphics[width=\linewidth]{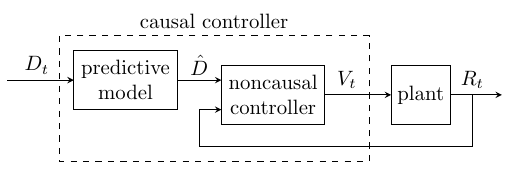}
	\caption{Block diagram depiction of proposed control scheme. The controller is composed of two components: a predictive model which forecasts the demand trajectory $\hat{D}$ from its current state $D_t$, and a noncausal controller which determines the control input $V_t$ by solving the necessary conditions for optimality \eqref{necc_cond_3} - \eqref{necc_cond_4} supposing $D = \hat{D}$.}
	\label{mpc_block_diag}
\end{figure}

Given a model for the demand and a method of producing forecasts, this scheme implicitly defines a feedback controller $\K$ which maps $(R_t,D_t)$ to a control input $V_t$. The controller $\K$ is causal, as it does not rely on future knowledge of the demand signal. While $\K$ is not in general optimal, it is interesting to note that if the demand is deterministic and the predictive model can predict the resulting trajectory $D$ exactly, then $\K$ produces the same control input $V_t$ as the optimal noncausal controller. In other words, for deterministic demands, $\K$ provides a causal implementation of the noncausal optimal control. For nondeterministic demands, $\K$ is not in general optimal, but unlike the noncausal optimal control, it can actually be implemented. The structure of this controller is further emphasized in Figure \ref{mpc_block_diag}.

\vspace{-2mm}

In this paper, we treat neither the problem of modeling and predicting the demand nor the problem of solving the necessary conditions for optimality, electing to leave these problems to future work. Rather, we take a simpler approach here: predict the demand trajectory using a model for which we already have explicit solutions. The model which we will use is that of a static demand signal
\begin{equation} \label{constant}
	\partial_t D_t ~=~ 0 ,
\end{equation}
the motivation being that since we already have explicit solutions for this case, the resulting MPC scheme is straightforward to implement and computationally attractive.

We point out that despite appearances, this is actually not too bad of a model. Note that we are not assuming that the demand is truly static. Rather, we are only using this model in the prediction step that anticipates the demand's next move. In the absence of any information about the behavior of the demand, predicting that the demand will stay where it is seems quite justifiable. Philosophically, by using this model, we are just saying that we cannot predict the demand's future.

We will show that even with the trivial model \eqref{constant}, the closed-loop performance of the resulting controller is quite reasonable. Because the state of the demand is updated with each timestep, the demand signal enters the control algorithm as a piecewise constant approximation. For continuous demand trajectories, the error in this approximation is first-order in the timestep $\Delta_t$, and thus as $\Delta_t \to 0$, this approximation converges to the true demand signal, and the controller reacts to changes in the demand in real time.

\subsection{Description of Controller}

In \cite{Emerick2023}, we showed that the optimal controller for problem \eqref{original} with a static demand signal $D_t = \bar{D}$ and an infinite time horizon $T = \infty$ is given by
\begin{equation} \label{static_controller}
	V_{t} ~=~ \frac{1}{\alpha}\big( M_{R_{t} \to \bar{D}} - \id \big) ,
\end{equation}
where $M_{R_t \to \bar{D}}$ is the optimal transport map taking $R_t$ to $\bar{D}$ and $\id$ is the identity map on $\Omega$.
The model predictive control algorithm based on this solution proceeds as follows:

\begin{algorithm}
\caption{}
\begin{algorithmic}[1]
	\For{\(t = 0, \, \Delta_t, \, 2\Delta_t, \, \ldots\)}
		\State Update the state of the demand to \(D_t\).
		\State For \(\tau\in[0, \Delta_t]\), apply the feedback control law
		\begin{equation} \label{const_mpc}
			V_{t+\tau} ~=~ \frac{1}{\alpha}\big( M_{R_{t+\tau} \to D_t} - \id \big) .
		\end{equation}
	\EndFor
\end{algorithmic}
\end{algorithm}

However, this algorithm -- as written above -- is difficult to implement, the main reason being that the term $M_{R_{t+\tau} \to D_t}$ is not very well behaved in these coordinates. Without much justification, let it suffice to say that we obtain a better-behaved and computationally more attractive solution by rewriting our system in different coordinates.

\subsection{From Eulerian to Lagrangian Coordinates}

Up until now, we have been describing our system in an \emph{Eulerian} framework. That is, we have been talking about how densities on the spatial domain $\Omega$ change with time. We can also describe our system in a \emph{Lagrangian} framework, which expresses the evolution of the system in terms of the trajectories of individual particles. We will briefly describe the relevant features of this correspondence here. For a full account, see \cite{Emerick2023}. The motivation for this coordinate change is that in many respects, particle trajectories are better behaved and easier to work with than time-varying densities. We point out that this is not a new idea -- this transformation is classical (see e.g. \cite{Cavagnari2022}) and essentially amounts to solution by the method of characteristics.

In the Lagrangian framework, we begin by assigning each particle an index $i$, and then represent the state of the resource with a map $Q$ which gives the time-varying positions of each particle. The value $Q_t(i)$ gives the position at time $t$ of the particle with index $i$. The map $Q$ and the resource distribution $R$ are related by the equation
\begin{equation} \label{pushforward}
	R_t = [Q_t]_\# \mu ,
\end{equation}
where $\#$ denotes the measure pushforward and $\mu$ is a fixed reference distribution describing the distribution of particle indices. The dynamics for our system can be expressed in terms of the map $Q$ as
\begin{equation} \label{Q_dynamics}
	\partial_t Q_t(i) ~=~ V_t \of Q_t(i) ~=:~ U_t(i) .
\end{equation}
Here, $V$ is our original vector field, $\of$ denotes function composition, and $U$ is the pullback of $V$ to the space of particle indices (i.e., $U(i)$ is simply the velocity of particle $i$). Notice that the dynamics of different particles are decoupled and are linear in the transformed input $U$.
The control \eqref{const_mpc} can now be written in Lagrangian coordinates as
\begin{align}
	U_{t+\tau} &~=~ V_{t+\tau} \of Q_{t+\tau} \nonumber \\
	&~=~ \frac{1}{\alpha} \big( M_{R_{t+\tau} \to D_t} \of Q_{t+\tau} - Q_{t+\tau} \big) \label{transformed_control} \\
	&~=:~ \frac{1}{\alpha} \big( \tilde{M}_{R_{t+\tau} \to D_t} - Q_{t+\tau} \big) , \nonumber
\end{align}
where $\tilde{M}$ is the pullback of the optimal transport map $M$ to the space of particle indices.
Notice that this control law is written in feedback form. However, feedback requires that the optimal transport map $\tilde{M}$ be continuously updated. Since computing $\tilde{M}$ is the bottleneck in this algorithm, we obtain a more computationally friendly solution by writing the control in open-loop form. To do this, we employ the following fact.
\begin{lem}
	Under the dynamics \eqref{Q_dynamics}, with control \eqref{transformed_control}, $Q$ evolves such that $\tilde{M}_{R_{t+\tau} \to D_t}$ remains constant. In particular,
	\begin{equation} \label{m_const}
		\tilde{M}_{R_{t+\tau} \to D_t} ~=~ \tilde{M}_{R_t \to D_t} .
	\end{equation}
\end{lem}

The proof of this lemma will be provided in a follow-up version of the paper.
Making the substitutions \eqref{transformed_control} and \eqref{m_const} in \eqref{Q_dynamics} and solving for $Q_{t+\tau}$ in terms of $Q_t$, we find that
\begin{equation} \label{Q_soln} 
	Q_{t+\tau} ~=~ (1-e^{-\tau/\alpha}) \, \tilde{M}_{R_t \to D_t} ~+~ e^{-\tau/\alpha} \, Q_t ,
\end{equation}
which we can differentiate to obtain the open-loop control
\begin{equation} \label{open_loop}
	U_{t+\tau} ~=~ \frac{1}{\alpha} e^{-\tau/\alpha} \, \big( \tilde{M}_{R_t \to D_t} - Q_t \big) .
\end{equation}

\subsection{Computational Algorithm}

We first discuss how we discretize our MPC scheme in space. This yields a continuous-time particle method which is well-suited for implementation on real-world swarms (as real-world swarms are themselves particle-based systems). Later, we will discuss how we discretize our scheme in time for the purposes of numerical simulation.

To discretize, we choose the set of particle indices to be a finite index set $I = \{ 1, 2, ..., N \}$, and choose the reference distribution as
\begin{equation}
	\mu ~=~ \frac{1}{N} \sum_{i=1}^N \delta_i
\end{equation}
for $i \in I$. Here, $\delta_i$ is the Dirac mass at $i$. The relation \eqref{pushforward} then tells us that the resource distribution is given by
\begin{equation} \label{discrete_resource}
	R_t ~=~ [Q_t]_\# \mu ~=~  \frac{1}{N} \sum_{i=1}^N \delta_{Q_t(i)} .
\end{equation}
In other words, the resource distribution is given by a collection of $N$ point particles indexed from $i = 1, ..., N$, each with position $Q_t(i)$ and mass $1/N$. The discrete-particle continuous-time MPC algorithm then proceeds as follows:

\begin{algorithm}[H]
\caption{}
\label{alg_mpc}
\begin{algorithmic}[1]
	\State Discretize $R_0$ and assign particles indices $i=1$ to $N$.
	\For{\(t = 0, \, \Delta_t, \, 2\Delta_t, \, \ldots\)}
		\State Update the state of the demand to \(D_t\).
		\State Compute the optimal transport map $\tilde{M}_{R_t \to D_t}$.
		\State For \(\tau\in[0, \Delta_t]\), apply the control \eqref{open_loop}.
	\EndFor
\end{algorithmic}
\end{algorithm}

Some comments on this control algorithm are in order. First, in a practical setting, the resource distribution is likely already discrete. However, if it is not, then it is not entirely obvious how best to approximate an arbitrary distribution with a discrete one. This is an issue that we will explore in future work. We will discuss how we treat this issue in the context of our numerical simulations later in Section \ref{simulations}.

Second, it is not obvious how best to compute (or approximate) the optimal transport map $\tilde{M}$, especially since optimal maps do not in general exist when $R_t$ is discrete. This is an active research area and goes by the name of \emph{discrete optimal transport}. In practice, $D_t$ is first discretized, then, either the Kantorovich problem or its dual is solved (approximated) numerically, and the ``barycentric map'' is computed from the solution to this problem. The barycenteric map approximates the optimal map in the sense that as the discretizations are refined, the barycenteric map converges to the optimal map for the continuous problem. See \cite{Peyre2019} for an in-depth discussion of these computational issues.

We now discuss how this numerical algorithm is discretized in time. We lose no generality here by assuming that the timestep of discretization is the same as the controller timestep $\Delta_t$, since we can always find intermediate states via \eqref{Q_soln}. We use $k$ for our discrete time variable, where $k = t/ \Delta_t$ or $t = k \Delta_t$. Thus, we simulate this MPC algorithm in discrete time using the dynamics
\begin{align}
	Q_{k+1} ~&=~ Q_k + \tilde{U}_k , \\
	\tilde{U}_k ~&:=~ \int_0^{\Delta_t} U_{t + \tau} \, d\tau \\
	~&=~ (1-e^{-\Delta_t / \alpha}) \big( \tilde{M}_{R_k \to D_k} - Q_k \big) .
\end{align}
This yields the update formula
\begin{equation} \label{eq_mpc_flow}
	\begin{split}
		Q_{k+1} ~&=~ Q_k + (1-e^{-\Delta_t / \alpha}) \big( \tilde{M}_{R_k \to D_k} - Q_k \big) \\
		~&=~ (1-e^{-\Delta_t / \alpha}) \, \tilde{M}_{R_k \to D_k} + e^{-\Delta_t / \alpha} \, Q_k .
	\end{split}
\end{equation}

\section{Simulations} \label{simulations}

In this section, we present simulation results applying the proposed MPC algorithm to various classes of demand signals. Continuing the discussion from the previous section, we first discretize the resource and demand distributions on a \(400\times 400\) grid on the unit square. We then generate particle representations for the resource and demand using inverse transform sampling. After this discretization, the resource and demand are approximated by discrete distributions (i.e. point clouds) of the form \eqref{discrete_resource}. The simulations that follow are implemented using Python, and are built on the Python package \emph{Optimal Transport Tools} (OTT) \cite{Cuturi2022}.

\begin{figure}[t]
	\centering
	\includegraphics[width=0.8\linewidth]{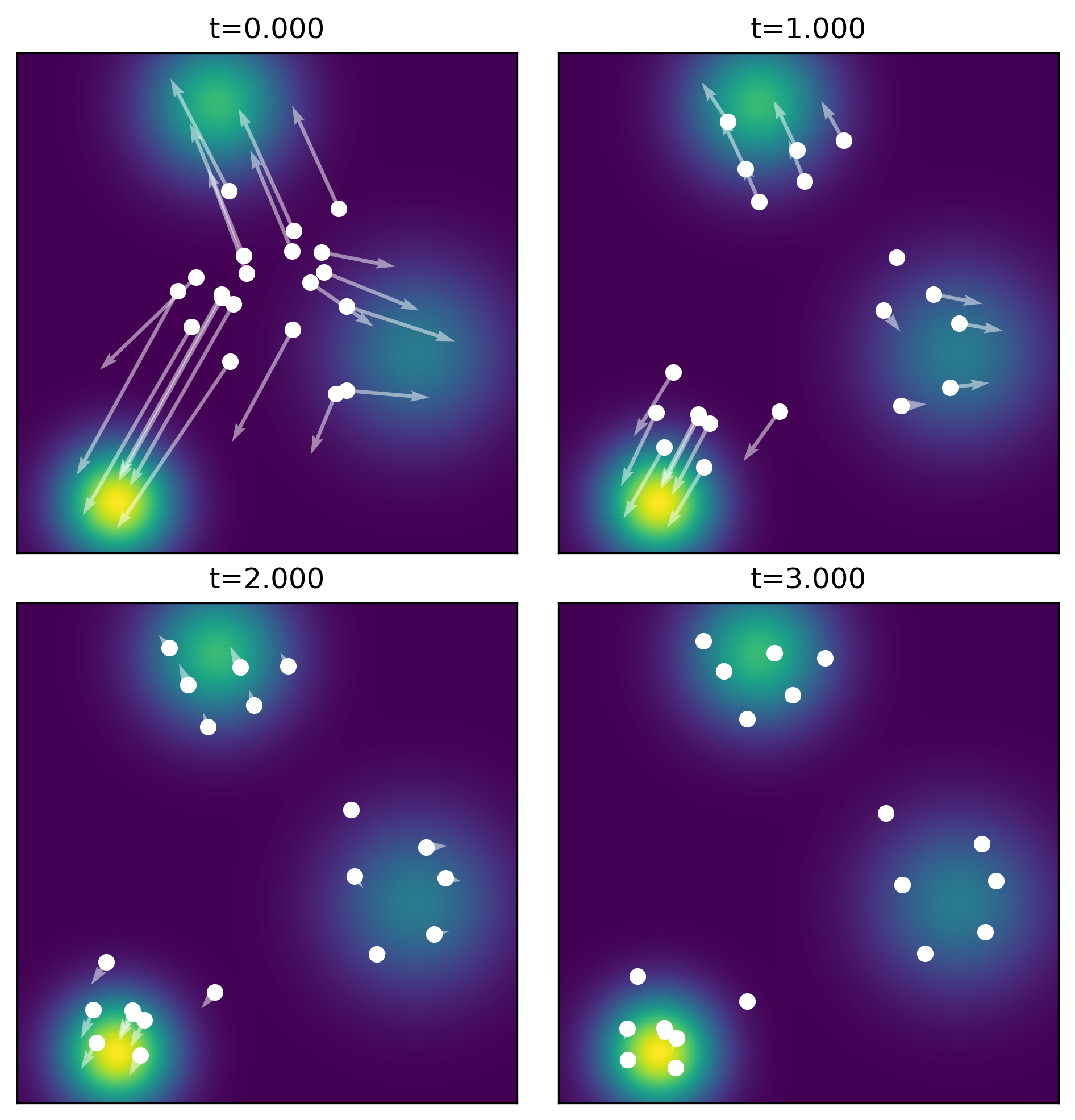}
	\caption{Timeseries for the static demand case.  Here, 20 resource particles in white track the demand, which is given by a Gaussian mixture.  The colored image represents the density of the demand distribution at the given time, where blue is less dense and yellow is more dense. The arrows attached to each particle show their respective assigned positions as determined by the optimal transport map. }
	\label{fig_static_ts}
\end{figure}

At \(t=0\), we start with a resource distribution of \(N\) particles.  At each timestep of the MPC algorithm~\ref{alg_mpc}, we first discretize the demand distribution using inverse transform sampling, and then find an approximation to the optimal transport map by solving the entropy-regularized dual Kantorovich problem using the Sinkhorn algorithm, which is implemented in OTT. Between MPC update times,  the points in the resource distribution are updated using equation~(\ref{eq_mpc_flow}).  Increasing the number of points in the demand distribution \(N_d\) or decreasing the entropic regularization weight \(\varepsilon\) both improve the numerical accuracy of the simulations. In the following simulations, we used \(N_d = 500\) and \(\varepsilon = 0.005\). In the following sections, we present simulation results for three different classes of demand signals.  Animations of each example can be viewed at  \url{https://tinyurl.com/2sk7c9wh}.  

\subsection{Static Demand}
We now consider a demand distribution that is static (i.e. constant in time).  In this case, the predictive model is exact. In the continuum version of this problem, the resource travels along the Wasserstein geodesic towards the demand, and the assigned locations of each resource particle remain constant. However, note that due to the discrete approximation, we observe small fluctuations in the assigned locations which settle down as time increases. Figure \ref{fig_static_ts} shows the behavior of the MPC scheme for a demand given by a Gaussian mixture. In this simulation, we used the parameters \(\alpha=0.1\) and \(\Delta_t = 0.1\).  At time \(t=0\), the resource particles are loosely scattered in the center.  As time progresses, they travel toward their assigned positions according to equation~(\ref{eq_mpc_flow}). We observe that the velocity decreases over time, as we should expect from this equation.

\subsection{Fading Demand}
\begin{figure}[t]
	\centering
	\includegraphics[width=0.8\linewidth]{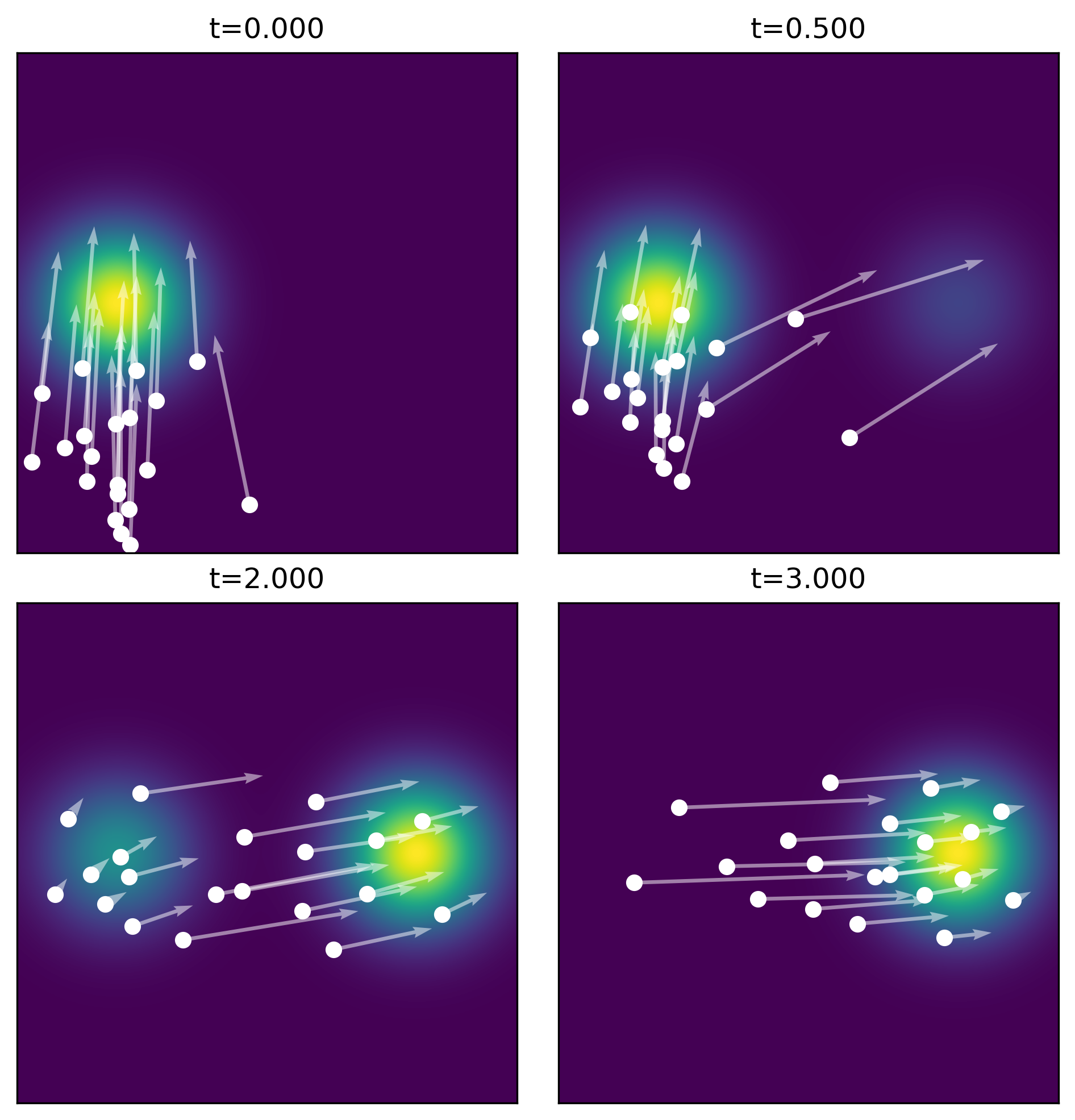}
	\caption{Timeseries for a time-varying demand. At \(t=0\) and \(t=3\), the demand is a Gaussian placed to the left or right side of the image respectively.  When \(0<t<3\), one fades out while the other fades in. }
	\label{fig_geodesic_ts}
\end{figure}

We now consider time-varying demand distributions. In Figure~\ref{fig_geodesic_ts}, all resource particles are initially assigned to positions on the left of the figure.  As the demand changes, each particle's assigned position changes until all of the particles are assigned to the right side at \(t=3\). Observe that the assigned positions and velocites of the resource particles react to changes in the demand in real time. In this simulation, we used the parameters \(\Delta_t = 0.05\) and \(\alpha = 1\).

\subsection{Constant-Velocity Demand}
\begin{figure}[ht]
	\centering
	\includegraphics[width=0.8\linewidth]{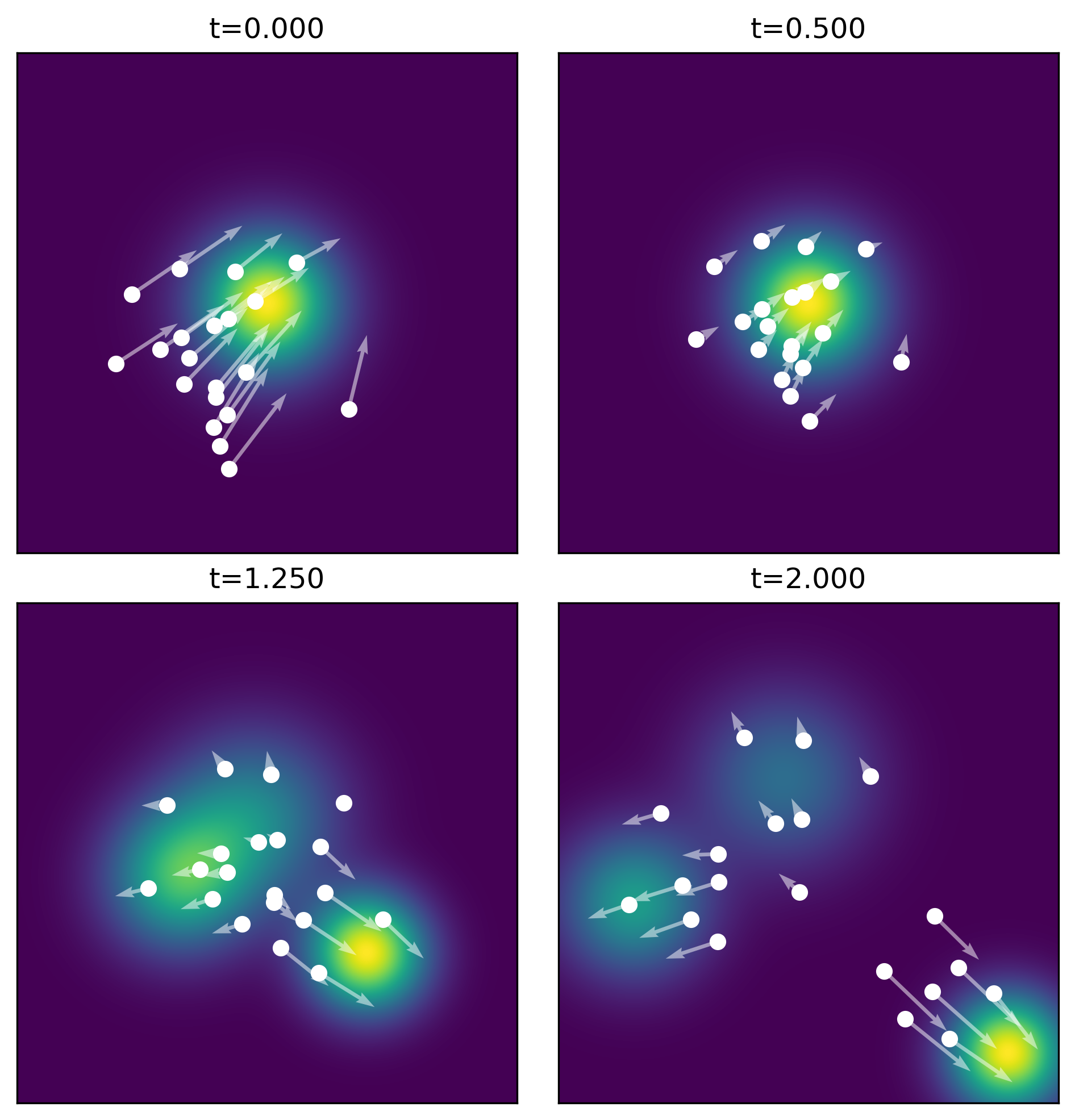}
	\caption{Timeseries for a time-varying demand distribution composed of three Gaussians. At first, all Gaussians start in the center. At \(t=0.5\), each begins to move away from the center at a constant velocity. }
	\label{fig_cv_ts}
\end{figure}

Figure~\ref{fig_cv_ts} shows a demand composed of three Gaussians which start in the center and then move outward with constant velocity beginning at $t = 0.5$.  In steady state, the particles lag behind their assigned locations at a constant distance which depends on the demand velocity and the parameter \(\alpha\).  As \(\alpha\) is increased, this distance increases.  Here, we used the parameters \(\Delta_t = 0.05\) and \(\alpha = 0.5\).

\section{Conclusion} \label{conclusion}

In this paper, we investigated a problem of optimal tracking control for swarms which are described in terms of their mass distributions. We described the noncausal structure of optimal solutions to this problem, and presented a model-predictive control scheme which uses these noncausal solutions together with a predictive demand model to achieve causal tracking control. We developed this model-predictive control scheme for the simple case of a static demand model and presented a computational algorithm based on particle methods and discrete optimal mass transport. We simulated the resulting controller for various classes of reference signals, and demonstrated that the proposed control algorithm is able to track time-varying references in real time. In future work, we plan to investigate more accurate demand models, computational methods for solving the resulting necessary conditions for optimality, and characterize the stability, performance, and robustness of the resulting controllers.


\bibliographystyle{ieeetr}
\bibliography{library}

\end{document}